\documentclass[10pt,conference,twocolumn]{IEEEtran}

\usepackage{cite}
\usepackage[T1]{fontenc}
\usepackage{graphicx}
\usepackage{amssymb}
\usepackage{amsmath}
\usepackage{amsthm}

\usepackage{paralist}
\usepackage{setspace}
\usepackage{microtype}
\usepackage{hyperref}
\usepackage{url}
\usepackage{balance}
\usepackage{subfigure}
\usepackage{xcolor}
\usepackage{fancyref}
\usepackage{framed}
\usepackage{booktabs}
\usepackage{nicefrac}
\usepackage{multirow}

\usepackage{algorithm}
\usepackage{algpseudocode}

\IEEEoverridecommandlockouts
\allowdisplaybreaks
\newtheorem{alg}{Algorithm}

\usepackage{amssymb}
\usepackage{amsfonts}
\usepackage{mathrsfs}
\usepackage{xspace}
\usepackage{bm}
\usepackage{upgreek}

\newcommand{\safemath}[2]{\newcommand{#1}{\ensuremath{#2}\xspace}}

\safemath{\bma}{\mathbf{a}}
\safemath{\bmb}{\mathbf{b}}
\safemath{\bmc}{\mathbf{c}}
\safemath{\bmd}{\mathbf{d}}
\safemath{\bme}{\mathbf{e}}
\safemath{\bmf}{\mathbf{f}}
\safemath{\bmg}{\mathbf{g}}
\safemath{\bmh}{\mathbf{h}}
\safemath{\bmi}{\mathbf{i}}
\safemath{\bmj}{\mathbf{j}}
\safemath{\bmk}{\mathbf{k}}
\safemath{\bml}{\mathbf{l}}
\safemath{\bmm}{\mathbf{m}}
\safemath{\bmn}{\mathbf{n}}
\safemath{\bmo}{\mathbf{o}}
\safemath{\bmp}{\mathbf{p}}
\safemath{\bmq}{\mathbf{q}}
\safemath{\bmr}{\mathbf{r}}
\safemath{\bms}{\mathbf{s}}
\safemath{\bmt}{\mathbf{t}}
\safemath{\bmu}{\mathbf{u}}
\safemath{\bmv}{\mathbf{v}}
\safemath{\bmw}{\mathbf{w}}
\safemath{\bmx}{\mathbf{x}}
\safemath{\bmy}{\mathbf{y}}
\safemath{\bmz}{\mathbf{z}}
\safemath{\bmzero}{\mathbf{0}}
\safemath{\bmone}{\mathbf{1}}

\bmdefine{\biad}{a}
\bmdefine{\bibd}{b}
\bmdefine{\bicd}{c}
\bmdefine{\bidd}{d}
\bmdefine{\bied}{e}
\bmdefine{\bifd}{f}
\bmdefine{\bigd}{g}
\bmdefine{\bihd}{h}
\bmdefine{\biid}{i}
\bmdefine{\bijd}{j}
\bmdefine{\bikd}{k}
\bmdefine{\bild}{l}
\bmdefine{\bimd}{m}
\bmdefine{\bind}{n}
\bmdefine{\biod}{o}
\bmdefine{\bipd}{p}
\bmdefine{\biqd}{q}
\bmdefine{\bird}{r}
\bmdefine{\bisd}{s}
\bmdefine{\bitd}{t}
\bmdefine{\biud}{u}
\bmdefine{\bivd}{v}
\bmdefine{\biwd}{w}
\bmdefine{\bixd}{x}
\bmdefine{\biyd}{y}
\bmdefine{\bizd}{z}

\bmdefine{\bixid}{\xi}
\bmdefine{\bilambdad}{\lambda}
\bmdefine{\bimud}{\mu}
\bmdefine{\bithetad}{\theta}
\bmdefine{\biphid}{\phi}
\bmdefine{\bideltad}{\delta}

\safemath{\bmia}{\biad}
\safemath{\bmib}{\bibd}
\safemath{\bmic}{\bicd}
\safemath{\bmid}{\bidd}
\safemath{\bmie}{\bied}
\safemath{\bmif}{\bifd}
\safemath{\bmig}{\bigd}
\safemath{\bmih}{\bihd}
\safemath{\bmii}{\biid}
\safemath{\bmij}{\bijd}
\safemath{\bmik}{\bikd}
\safemath{\bmil}{\bild}
\safemath{\bmim}{\bimd}
\safemath{\bmin}{\bind}
\safemath{\bmio}{\biod}
\safemath{\bmip}{\bipd}
\safemath{\bmiq}{\biqd}
\safemath{\bmir}{\bird}
\safemath{\bmis}{\bisd}
\safemath{\bmit}{\bitd}
\safemath{\bmiu}{\biud}
\safemath{\bmiv}{\bivd}
\safemath{\bmiw}{\biwd}
\safemath{\bmix}{\bixd}
\safemath{\bmiy}{\biyd}
\safemath{\bmiz}{\bizd}

\safemath{\bmxi}{\bixid}
\safemath{\bmlambda}{\bilambdad}
\safemath{\bmmu}{\bimud}
\safemath{\bmtheta}{\bithetad}
\safemath{\bmphi}{\biphid}
\safemath{\bmdelta}{\bideltad}

\safemath{\bA}{\mathbf{A}}
\safemath{\bB}{\mathbf{B}}
\safemath{\bC}{\mathbf{C}}
\safemath{\bD}{\mathbf{D}}
\safemath{\bE}{\mathbf{E}}
\safemath{\bF}{\mathbf{F}}
\safemath{\bG}{\mathbf{G}}
\safemath{\bH}{\mathbf{H}}
\safemath{\bI}{\mathbf{I}}
\safemath{\bJ}{\mathbf{J}}
\safemath{\bK}{\mathbf{K}}
\safemath{\bL}{\mathbf{L}}
\safemath{\bM}{\mathbf{M}}
\safemath{\bN}{\mathbf{N}}
\safemath{\bO}{\mathbf{O}}
\safemath{\bP}{\mathbf{P}}
\safemath{\bQ}{\mathbf{Q}}
\safemath{\bR}{\mathbf{R}}
\safemath{\bS}{\mathbf{S}}
\safemath{\bT}{\mathbf{T}}
\safemath{\bU}{\mathbf{U}}
\safemath{\bV}{\mathbf{V}}
\safemath{\bW}{\mathbf{W}}
\safemath{\bX}{\mathbf{X}}
\safemath{\bY}{\mathbf{Y}}
\safemath{\bZ}{\mathbf{Z}}

\safemath{\bZero}{\mathbf{0}}
\safemath{\bOne}{\mathbf{1}}
\safemath{\bDelta}{\mathbf{\Delta}}
\safemath{\bLambda}{\mathbf{\UpLambda}}
\safemath{\bPhi}{\mathbf{\Upphi}}
\safemath{\bSigma}{\mathbf{\Upsigma}}
\safemath{\bOmega}{\mathbf{\Upomega}}
\safemath{\bTheta}{\mathbf{\Uptheta}}

\bmdefine{\biAd}{A}
\bmdefine{\biBd}{B}
\bmdefine{\biCd}{C}
\bmdefine{\biDd}{D}
\bmdefine{\biEd}{E}
\bmdefine{\biFd}{F}
\bmdefine{\biGd}{G}
\bmdefine{\biHd}{H}
\bmdefine{\biId}{I}
\bmdefine{\biJd}{J}
\bmdefine{\biKd}{K}
\bmdefine{\biLd}{L}
\bmdefine{\biMd}{M}
\bmdefine{\biOd}{N}
\bmdefine{\biPd}{O}
\bmdefine{\biQd}{P}
\bmdefine{\biRd}{R}
\bmdefine{\biSd}{S}
\bmdefine{\biTd}{T}
\bmdefine{\biUd}{U}
\bmdefine{\biVd}{V}
\bmdefine{\biWd}{W}
\bmdefine{\biXd}{X}
\bmdefine{\biYd}{Y}
\bmdefine{\biZd}{Z}

\bmdefine{\biDelta}{\Delta}
\bmdefine{\biLambda}{\Lambda}
\bmdefine{\biPhi}{\Phi}
\bmdefine{\biSigma}{\Sigma}
\bmdefine{\biOmega}{\Omega}
\bmdefine{\biTheta}{\Theta}

\safemath{\bimA}{\biAd}
\safemath{\bimB}{\biBd}
\safemath{\bimC}{\biCd}
\safemath{\bimD}{\biDd}
\safemath{\bimE}{\biEd}
\safemath{\bimF}{\biFd}
\safemath{\bimG}{\biGd}
\safemath{\bimH}{\biHd}
\safemath{\bimI}{\biId}
\safemath{\bimJ}{\biJd}
\safemath{\bimK}{\biKd}
\safemath{\bimL}{\biLd}
\safemath{\bimM}{\biMd}
\safemath{\bimN}{\biNd}
\safemath{\bimO}{\biOd}
\safemath{\bimP}{\biPd}
\safemath{\bimQ}{\biQd}
\safemath{\bimR}{\biRd}
\safemath{\bimS}{\biSd}
\safemath{\bimT}{\biTd}
\safemath{\bimU}{\biUd}
\safemath{\bimV}{\biVd}
\safemath{\bimW}{\biWd}
\safemath{\bimX}{\biXd}
\safemath{\bimY}{\biYd}
\safemath{\bimZ}{\biZd}

\safemath{\bimDelta}{\biDelta}
\safemath{\bimLambda}{\biLambda}
\safemath{\bimPhi}{\biPhi}
\safemath{\bimSigma}{\biSigma}
\safemath{\bimOmega}{\biOmega}
\safemath{\bimTheta}{\biTheta}

\safemath{\setA}{\mathcal{A}}
\safemath{\setB}{\mathcal{B}}
\safemath{\setC}{\mathcal{C}}
\safemath{\setD}{\mathcal{D}}
\safemath{\setE}{\mathcal{E}}
\safemath{\setF}{\mathcal{F}}
\safemath{\setG}{\mathcal{G}}
\safemath{\setH}{\mathcal{H}}
\safemath{\setI}{\mathcal{I}}
\safemath{\setJ}{\mathcal{J}}
\safemath{\setK}{\mathcal{K}}
\safemath{\setL}{\mathcal{L}}
\safemath{\setM}{\mathcal{M}}
\safemath{\setN}{\mathcal{N}}
\safemath{\setO}{\mathcal{O}}
\safemath{\setP}{\mathcal{P}}
\safemath{\setQ}{\mathcal{Q}}
\safemath{\setR}{\mathcal{R}}
\safemath{\setS}{\mathcal{S}}
\safemath{\setT}{\mathcal{T}}
\safemath{\setU}{\mathcal{U}}
\safemath{\setV}{\mathcal{V}}
\safemath{\setW}{\mathcal{W}}
\safemath{\setX}{\mathcal{X}}
\safemath{\setY}{\mathcal{Y}}
\safemath{\setZ}{\mathcal{Z}}
\safemath{\emptySet}{\varnothing}

\safemath{\colA}{\mathscr{A}}
\safemath{\colB}{\mathscr{B}}
\safemath{\colC}{\mathscr{C}}
\safemath{\colD}{\mathscr{D}}
\safemath{\colE}{\mathscr{E}}
\safemath{\colF}{\mathscr{F}}
\safemath{\colG}{\mathscr{G}}
\safemath{\colH}{\mathscr{H}}
\safemath{\colI}{\mathscr{I}}
\safemath{\colJ}{\mathscr{J}}
\safemath{\colK}{\mathscr{K}}
\safemath{\colL}{\mathscr{L}}
\safemath{\colM}{\mathscr{M}}
\safemath{\colN}{\mathscr{N}}
\safemath{\colO}{\mathscr{O}}
\safemath{\colP}{\mathscr{P}}
\safemath{\colQ}{\mathscr{Q}}
\safemath{\colR}{\mathscr{R}}
\safemath{\colS}{\mathscr{S}}
\safemath{\colT}{\mathscr{T}}
\safemath{\colU}{\mathscr{U}}
\safemath{\colV}{\mathscr{V}}
\safemath{\colW}{\mathscr{W}}
\safemath{\colX}{\mathscr{X}}
\safemath{\colY}{\mathscr{Y}}
\safemath{\colZ}{\mathscr{Z}}

\safemath{\opA}{\mathbb{A}}
\safemath{\opB}{\mathbb{B}}
\safemath{\opC}{\mathbb{C}}
\safemath{\opD}{\mathbb{D}}
\safemath{\opE}{\mathbb{E}}
\safemath{\opF}{\mathbb{F}}
\safemath{\opG}{\mathbb{G}}
\safemath{\opH}{\mathbb{H}}
\safemath{\opI}{\mathbb{I}}
\safemath{\opJ}{\mathbb{J}}
\safemath{\opK}{\mathbb{K}}
\safemath{\opL}{\mathbb{L}}
\safemath{\opM}{\mathbb{M}}
\safemath{\opN}{\mathbb{N}}
\safemath{\opO}{\mathbb{O}}
\safemath{\opP}{\mathbb{P}}
\safemath{\opQ}{\mathbb{Q}}
\safemath{\opR}{\mathbb{R}}
\safemath{\opS}{\mathbb{S}}
\safemath{\opT}{\mathbb{T}}
\safemath{\opU}{\mathbb{U}}
\safemath{\opV}{\mathbb{V}}
\safemath{\opW}{\mathbb{W}}
\safemath{\opX}{\mathbb{X}}
\safemath{\opY}{\mathbb{Y}}
\safemath{\opZ}{\mathbb{Z}}
\safemath{\opZero}{\mathbb{O}}
\safemath{\identityop}{\opI}

\safemath{\veca}{\bma}
\safemath{\vecb}{\bmb}
\safemath{\vecc}{\bmc}
\safemath{\vecd}{\bmd}
\safemath{\vece}{\bme}
\safemath{\vecf}{\bmf}
\safemath{\vecg}{\bmg}
\safemath{\vech}{\bmh}
\safemath{\veci}{\bmi}
\safemath{\vecj}{\bmj}
\safemath{\veck}{\bmk}
\safemath{\vecl}{\bml}
\safemath{\vecm}{\bmm}
\safemath{\vecn}{\bmn}
\safemath{\veco}{\bmo}
\safemath{\vecp}{\bmp}
\safemath{\vecq}{\bmq}
\safemath{\vecr}{\bmr}
\safemath{\vecs}{\bms}
\safemath{\vect}{\bmt}
\safemath{\vecu}{\bmu}
\safemath{\vecv}{\bmv}
\safemath{\vecw}{\bmw}
\safemath{\vecx}{\bmx}
\safemath{\vecy}{\bmy}
\safemath{\vecz}{\bmz}

\safemath{\veczero}{\bmzero}
\safemath{\vecone}{\bmone}
\safemath{\vecxi}{\bmxi}
\safemath{\veclambda}{\bmlambda}
\safemath{\vecmu}{\bmmu}
\safemath{\vectheta}{\bmtheta}
\safemath{\vecphi}{\bmphi}
\safemath{\vecdelta}{\bmdelta}

\safemath{\matA}{\bA}
\safemath{\matB}{\bB}
\safemath{\matC}{\bC}
\safemath{\matD}{\bD}
\safemath{\matE}{\bE}
\safemath{\matF}{\bF}
\safemath{\matG}{\bG}
\safemath{\matH}{\bH}
\safemath{\matI}{\bI}
\safemath{\matJ}{\bJ}
\safemath{\matK}{\bK}
\safemath{\matL}{\bL}
\safemath{\matM}{\bM}
\safemath{\matN}{\bN}
\safemath{\matO}{\bO}
\safemath{\matP}{\bP}
\safemath{\matQ}{\bQ}
\safemath{\matR}{\bR}
\safemath{\matS}{\bS}
\safemath{\matT}{\bT}
\safemath{\matU}{\bU}
\safemath{\matV}{\bV}
\safemath{\matW}{\bW}
\safemath{\matX}{\bX}
\safemath{\matY}{\bY}
\safemath{\matZ}{\bZ}
\safemath{\matzero}{\bmzero}

\safemath{\matDelta}{\bDelta}
\safemath{\matLambda}{\bLambda}
\safemath{\matPhi}{\bPhi}
\safemath{\matSigma}{\bSigma}
\safemath{\matOmega}{\bOmega}
\safemath{\matTheta}{\bTheta}

\safemath{\matidentity}{\matI}
\safemath{\matone}{\matO}

\safemath{\rnda}{A}
\safemath{\rndb}{B}
\safemath{\rndc}{C}
\safemath{\rndd}{D}
\safemath{\rnde}{E}
\safemath{\rndf}{F}
\safemath{\rndg}{G}
\safemath{\rndh}{H}
\safemath{\rndi}{I}
\safemath{\rndj}{J}
\safemath{\rndk}{K}
\safemath{\rndl}{L}
\safemath{\rndm}{M}
\safemath{\rndn}{N}
\safemath{\rndo}{O}
\safemath{\rndp}{P}
\safemath{\rndq}{Q}
\safemath{\rndr}{R}
\safemath{\rnds}{S}
\safemath{\rndt}{T}
\safemath{\rndu}{U}
\safemath{\rndv}{V}
\safemath{\rndw}{W}
\safemath{\rndx}{X}
\safemath{\rndy}{Y}
\safemath{\rndz}{Z}

\safemath{\rveca}{\bimA}
\safemath{\rvecb}{\bimB}
\safemath{\rvecc}{\bimC}
\safemath{\rvecd}{\bimD}
\safemath{\rvece}{\bimE}
\safemath{\rvecf}{\bimF}
\safemath{\rvecg}{\bimG}
\safemath{\rvech}{\bimH}
\safemath{\rveci}{\bimI}
\safemath{\rvecj}{\bimJ}
\safemath{\rveck}{\bimK}
\safemath{\rvecl}{\bimL}
\safemath{\rvecm}{\bimM}
\safemath{\rvecn}{\bimN}
\safemath{\rveco}{\bomO}
\safemath{\rvecp}{\bimP}
\safemath{\rvecq}{\bimQ}
\safemath{\rvecr}{\bimR}
\safemath{\rvecs}{\bimS}
\safemath{\rvect}{\bimT}
\safemath{\rvecu}{\bimU}
\safemath{\rvecv}{\bimV}
\safemath{\rvecw}{\bimW}
\safemath{\rvecx}{\bimX}
\safemath{\rvecy}{\bimY}
\safemath{\rvecz}{\bimZ}

\safemath{\rvecxi}{\bmxi}
\safemath{\rveclambda}{\bmlambda}
\safemath{\rvecmu}{\bmmu}
\safemath{\rvectheta}{\bmtheta}
\safemath{\rvecphi}{\bmphi}

\safemath{\rmatA}{\bimA}
\safemath{\rmatB}{\bimB}
\safemath{\rmatC}{\bimC}
\safemath{\rmatD}{\bimD}
\safemath{\rmatE}{\bimE}
\safemath{\rmatF}{\bimF}
\safemath{\rmatG}{\bimG}
\safemath{\rmatH}{\bimH}
\safemath{\rmatI}{\bimI}
\safemath{\rmatJ}{\bimJ}
\safemath{\rmatK}{\bimK}
\safemath{\rmatL}{\bimL}
\safemath{\rmatM}{\bimM}
\safemath{\rmatN}{\bimN}
\safemath{\rmatO}{\bimO}
\safemath{\rmatP}{\bimP}
\safemath{\rmatQ}{\bimQ}
\safemath{\rmatR}{\bimR}
\safemath{\rmatS}{\bimS}
\safemath{\rmatT}{\bimT}
\safemath{\rmatU}{\bimU}
\safemath{\rmatV}{\bimV}
\safemath{\rmatW}{\bimW}
\safemath{\rmatX}{\bimX}
\safemath{\rmatY}{\bimY}
\safemath{\rmatZ}{\bimZ}

\safemath{\rmatDelta}{\bimDelta}
\safemath{\rmatLambda}{\bimLambda}
\safemath{\rmatPhi}{\bimPhi}
\safemath{\rmatSigma}{\bimSigma}
\safemath{\rmatOmega}{\bimOmega}
\safemath{\rmatTheta}{\bimTheta}

\usepackage{amssymb}
\usepackage{amsfonts}
\usepackage{mathrsfs}
\usepackage{xspace}
\usepackage{bm}
\usepackage{fancyref}
\usepackage{textcomp}

\usepackage{multirow}
\usepackage{stmaryrd}

\newenvironment{textbmatrix}{	\setlength{\arraycolsep}{2.5pt}%
								\big[\begin{matrix}}{\end{matrix}\big]%
								\raisebox{0.08ex}{\vphantom{M}}}

\def\be{\begin{equation}}
\def\ee{\end{equation}}
\def\een{\nonumber \end{equation}}
\def\mat{\begin{bmatrix}}
\def\emat{\end{bmatrix}}
\def\btm{\begin{textbmatrix}}
\def\etm{\end{textbmatrix}}

\def\ba#1\ea{\begin{align}#1\end{align}}
\def\bas#1\eas{\begin{align*}#1\end{align*}}
\def\bs#1\es{\begin{split}#1\end{split}} 
\def\bg#1\eg{\begin{gather}#1\end{gather}}
\def\bml#1\eml{\begin{multline}#1\end{multline}}
\def\bi#1\ei{\begin{itemize}#1\end{itemize}}

\newcommand{\lefto}{\mathopen{}\left}

\DeclareMathOperator{\tr}{tr}				% trace
\DeclareMathOperator*{\argmin}{arg\;min}		% arg min
\DeclareMathOperator{\Exop}{\opE}			% expectation operator
\newcommand{\Ex}[2]{\ensuremath{\Exop_{#1}\lefto[#2\right]}} 	% expectation
\newcommand{\vecnorm}[1]{\lefto\lVert#1\right\rVert}		% vector norm
\safemath{\dirac}{\delta}					% Dirac delta
\safemath{\krond}{\dirac}					% Kronecker delta
\safemath{\upto}{\uparrow}
\safemath{\downto}{\downarrow}
\safemath{\iu}{j}							% imaginary unit
\safemath{\ev}{\lambda}						% eigenvalue
\safemath{\hilseqspace}{l^{2}}				% Hilbert sequence space
\newcommand{\banachfunspace}[1]{\setL^{#1}}	% Banach function space
\safemath{\hilfunspace}{\banachfunspace{2}}	% Hilbert function space
\safemath{\SNR}{\textsf{SNR}} 				% signal to noise ratio
\safemath{\PAR}{\textsf{PAR}} 				% signal to noise ratio
\safemath{\No}{N_0}							% noise spectral density
\safemath{\Es}{E_s}							% energy per symbol
\safemath{\Eb}{E_b}							% energy per bit
\safemath{\EbNo}{\frac{\Eb}{\No}}
\safemath{\EsNo}{\frac{\Es}{\No}}

\DeclareMathOperator{\CHop}{\ensuremath{\opH}} % channel operator
\safemath{\tvir}{\rndh_{\CHop}}				% time-varying impulse response
\safemath{\tvtf}{\rndl_{\CHop}}				% 	-''- transfer function
\safemath{\spf}{\rnds_{\CHop}}				% spreading function
\safemath{\bff}{H_{\CHop}}					% bi-freuqency function

\safemath{\ircf}{r_{h}}						% impulse response correlation fn.
\safemath{\tftvcf}{r_{s}}					% scattering function
\safemath{\tfcf}{r_{l}}						% time-frequency correlation fn.
\safemath{\bfcf}{r_{H}}						% bi-frequency correlation fn.

\safemath{\tcorr}{c_h}						% time-correlation function
\safemath{\scf}{c_{s}}						% spreading function
\safemath{\tfcorr}{c_{l}}					% transfer-function correlation
\safemath{\fcorr}{c_{H}}						% frequency-correlation function

\safemath{\mi}{I}							% mutual information
\safemath{\capacity}{C}						% capacity

\safemath{\normal}{\mathcal{N}}			% normal distribution
\safemath{\jpg}{\mathcal{CN}}			% jointly proper Gaussian
\safemath{\mchain}{\leftrightarrow}		% Markov chain
\safemath{\dB}{\,\mathrm{dB}}
\safemath{\dBm}{\,\mathrm{dBm}}
\safemath{\Hz}{\,\mathrm{Hz}}
\safemath{\kHz}{\,\mathrm{kHz}}
\safemath{\MHz}{\,\mathrm{MHz}}
\safemath{\GHz}{\,\mathrm{GHz}}
\safemath{\s}{\,\mathrm{s}}
\safemath{\ms}{\,\mathrm{ms}}
\safemath{\mus}{\,\mathrm{\text{\textmu}s}}
\safemath{\ns}{\,\mathrm{ns}}
\safemath{\ps}{\,\mathrm{ps}}
\safemath{\meter}{\,\mathrm{m}}
\safemath{\mm}{\,\mathrm{mm}}
\safemath{\cm}{\,\mathrm{cm}}
\safemath{\m}{\,\mathrm{m}}
\safemath{\W}{\,\mathrm{W}}
\safemath{\mW}{\, \mathrm{mW}}
\safemath{\J}{\,\mathrm{J}}
\safemath{\K}{\,\mathrm{K}}
\safemath{\bit}{\,\mathrm{bit}}
\safemath{\nat}{\,\mathrm{nat}}

\safemath{\define}{\triangleq}			% definition

\safemath{\equivalent}{\sim}
\safemath{\distas}{\sim}					% distributed according to
\safemath{\sdiff}{\Delta}				% symmetric set difference

\safemath{\reals}{\mathbb{R}}
\safemath{\positivereals}{\reals_{+}}
\safemath{\integers}{\mathbb{Z}}
\safemath{\posint}{\integers_{+}}
\safemath{\naturals}{\mathbb{N}}
\safemath{\posnaturals}{\naturals_{+}}
\safemath{\complexset}{\mathbb{C}}
\safemath{\rationals}{\mathbb{Q}}

\newcommand*{\fancyrefapplabelprefix}{app}		% Appendix
\newcommand*{\fancyrefthmlabelprefix}{thm}		% Theorem
\newcommand*{\fancyreflemlabelprefix}{lem}		% Lemma
\newcommand*{\fancyrefcorlabelprefix}{cor}		% Corollary
\newcommand*{\fancyrefdeflabelprefix}{def}		% Definition
\newcommand*{\fancyrefproplabelprefix}{prop}	% Proposition
\newcommand*{\fancyrefobslabelprefix}{obs}		% Observation 
\newcommand*{\fancyrefalglabelprefix}{alg}		% Algorithm
\newcommand*{\fancyrefasmlabelprefix}{asm}	    % Assumption

\frefformat{vario}{\fancyrefseclabelprefix}{Section~#1}
\frefformat{vario}{\fancyrefthmlabelprefix}{Theorem~#1}
\frefformat{vario}{\fancyreflemlabelprefix}{Lemma~#1}
\frefformat{vario}{\fancyrefcorlabelprefix}{Corollary~#1}
\frefformat{vario}{\fancyrefdeflabelprefix}{Definition~#1}
\frefformat{vario}{\fancyrefobslabelprefix}{Observation~#1}
\frefformat{vario}{\fancyrefasmlabelprefix}{Assumption~#1}
\frefformat{vario}{\fancyreffiglabelprefix}{Fig.~#1}
\frefformat{vario}{\fancyrefapplabelprefix}{Appendix~#1} 
\frefformat{vario}{\fancyrefproplabelprefix}{Proposition~#1}
\frefformat{vario}{\fancyrefalglabelprefix}{Algorithm~#1}
\frefformat{vario}{\fancyrefeqlabelprefix}{(#1)}

\newtheorem{thm}{Theorem}
\newtheorem{defi}{Definition}
\newtheorem{lem}[thm]{Lemma} % Turned off lemma numbering

\safemath{\dictab}{[\,\dicta\,\,\dictb\,]}

\safemath{\ysig}{\bmy}
\safemath{\ysighat}{\hat{\ysig}}
\safemath{\ysigdim}{M}
\safemath{\xsig}{\bmx}
\safemath{\xsigdim}{N}
\safemath{\nx}{n_x}
\safemath{\zsig}{\bmz}
\safemath{\zsigdim}{\ysigdim}
\safemath{\rsig}{\bmr}
\safemath{\Adict}{\bA}
\safemath{\Adicttilde}{\widetilde{\Adict}}
\safemath{\Adictdim}{\outputdim\times\xsigdim}
\safemath{\avec}{\bma}
\safemath{\avectilde}{\tilde{\avec}}
\safemath{\Bdict}{\bB}
\safemath{\Bdicttilde}{\widetilde{\Bdict}}
\safemath{\Cdict}{\bC}
\safemath{\cvec}{\bmc}
\safemath{\Ddict}{\bD}
\safemath{\Ddictdim}{\ysigdim\times\xsigdim}
\safemath{\dvec}{\bmd}
\safemath{\Ddicttilde}{\widetilde{\bD}}
\safemath{\Bonb}{\bB}
\safemath{\bvec}{\bmb}
\safemath{\Bonbdim}{\ysigdim\times\ysigdim}
\safemath{\noise}{\bmn}
\safemath{\noisedim}{\ysigim}
\safemath{\err}{\bme}
\safemath{\errdim}{\ysigdim}
\safemath{\errset}{\setE}
\safemath{\nerr}{n_e}
\safemath{\delop}{\bP_\errset}
\safemath{\delopc}{\bP_{{\errset}^c}}

\safemath{\cplxi}{\imath}
\safemath{\cplxj}{\jmath}
\safemath{\dict}{\matD}
\safemath{\inputdim}{N}		% number of columns of dictionary D
\safemath{\outputdim}{M}		%number of rows of dictionary D
\safemath{\sparsity}{S}	%sparsity
\safemath{\inputdimA}{{N_a}}	%total number of elements in dictionary A
\safemath{\inputdimB}{{N_b}}	%total number of elements in dictionary B
\safemath{\elemA}{{n_a}}	%number of elements chosen from dictionary A
\safemath{\elemB}{{n_b}}	%number of elements chosen from dictionary B
\safemath{\resA}{\matR_a}	%restriction map to elements of dictionary A
\safemath{\resB}{\matR_b}	%restriction map to elements of dictionary B
\safemath{\subD}{\matS} %subdictionary
\safemath{\subA}{\matS_a} %subdictionary part of A
\safemath{\subB}{\matS_b} %subdictionary part of B
\safemath{\dicta}{\matA} 	% first subdictionary
\safemath{\dictb}{\matB} 	% second subdictionary
\safemath{\hollowS}{H}
\safemath{\hollowA}{H_a}
\safemath{\hollowB}{H_b}
\safemath{\cross}{Z}
\safemath{\coh}{\mu_d}			% coherence dictionary
\safemath{\coha}{\mu_a}			% coherence first subdictionary
\safemath{\cohb}{\mu_b}			% coherence second subdictionary
\safemath{\mubs}{\nu}	%block sub-coherence
\safemath{\cohm}{\mu_m} %mutual coherence
\safemath{\dictset}{\setD}	% set of dictionaries
\safemath{\dictsetp}{\dictset(\coh,\coha,\cohb)}	% set of dictionaries parametrized
\safemath{\dictsetgen}{\dictset_\text{gen}}
\safemath{\dictsetgenp}{\dictsetgen(\coh)}
\safemath{\dictsetonb}{\dictset_\text{onb}}
\safemath{\dictsetonbp}{\dictsetonb(\coh)}

\safemath{\leftside}{U}
\safemath{\rightsideA}{R_a}
\safemath{\rightsideB}{R_b}

\safemath{\indexS}{\setI_S} %set of indices participating in sub-dictionary S

\safemath{\na}{n_a}			% cardinality of set of linearly independent columns of first dictionary
\safemath{\nb}{n_b}			% cardinality of set of linearly independent columns of second dictionary
\safemath{\coeffa}{p_i}	%coefficients for columns of A
\safemath{\coeffb}{q_j}	%coefficients for columns of B
\safemath{\seta}{\setP}		% set of linearly independent columns of A
\safemath{\setb}{\setQ}     % set of linearly independent columns of B
\safemath{\setw}{\setW}	%set of n largest elements of w
\safemath{\setz}{\setZ}	%set of L-n largest elements of z
\safemath{\cola}{\veca}		% generic element of the dictionary A
\safemath{\colb}{\vecb}		% generic element of the dictionary B
\safemath{\cold}{\vecd}		% generic element of the dictionary D
\safemath{\inputvec}{\vecx} 	%coefficient vector (input)
\safemath{\error}{\vece}	%error vector
\safemath{\noiseout}{\vecz} 	%noisy output vector
\safemath{\inputvecel}{x}
\safemath{\inputveca}{\vecx_a}
\safemath{\inputvecb}{\vecx_b}
\safemath{\outputvec}{\vecy}	%output of Dictionary
\safemath{\lambdamin}{\lambda_{\mathrm{min}}}
\safemath{\elltwo}{\ell_2}
\safemath{\ellone}{\ell_1}
\safemath{\ellzero}{\ell_0}
\safemath{\ellinf}{\ell_\infty}
\safemath{\ellinftilde}{\ell_{\widetilde\infty}}
\safemath{\licard}{Z(\coh,\coha,\cohb)}
\safemath{\xsol}{\hat{x}}
\safemath{\xbord}{x_b}		%Solution at the border
\safemath{\xstat}{x_s}		%Solution stationary in l0 prob
\safemath{\xstatLone}{\tilde{x}_s}
\safemath{\order}{\mathcal{O}} %order notation (big O)
\safemath{\scales}{\Theta} %scales as
\safemath{\ones}{\mathbf{1}} %all ones matrix
\safemath{\zeroes}{\mathbf{0}} %all zeroes matrix
\safemath{\thlone}{\kappa(\coh,\cohb)} %treshold l1 problem
\safemath{\constoneA}{\delta} %constant in l1 theorem to save space
\safemath{\constoneB}{\epsilon} %constant in l1 theorem to save space
\safemath{\nlarge}{L}				   %num large elements
\safemath{\sumlarge}{S_\nlarge}
\safemath{\maxlarger}{P_\nlarge}	   % maximum in Gribonval and Nielsen
\safemath{\Pzero}{\textrm{P0}}	
\safemath{\Pone}{\textrm{P1}}
\safemath{\vecfir}{\vecw}			 % \vecv element of the kernel of the dictionary, \vecv=[\vecfir \vecsec]
\safemath{\vecsec}{\vecz}
\safemath{\elvecfir}{w}              % element of vecfir
\safemath{\elvecsec}{z}				 % element of vecsec
\safemath{\nlargefir}{n}
\safemath{\normout}{\gamma}
\safemath{\auxfun}{h}
\safemath{\supp}{\textrm{supp}}%support

\safemath{\indexa}{\ell}
\safemath{\indexb}{r}
\safemath{\indexc}{i}
\safemath{\indexd}{j}

\safemath{\project}{P}%projector

\makeatletter
\def\@IEEEinterspaceratioM{0.265}
\def\@IEEEinterspaceMINratioM{0.1651}
\def\@IEEEinterspaceMAXratioM{0.38}

\def\@IEEEinterspaceratioB{0.31}
\def\@IEEEinterspaceMINratioB{0.19}
\def\@IEEEinterspaceMAXratioB{0.38}
\@IEEEtunefonts
\makeatother
\begin{document}
% Title.
% ------

%
% Single address.
% ---------------
\title{Finite-Alphabet Wiener Filter Precoding \\ for mmWave Massive MU-MIMO Systems}
\author{\IEEEauthorblockN{Oscar Casta\~neda$^1$, Sven Jacobsson$^{2,3}$, Giuseppe Durisi$^3$, Tom Goldstein$^4$, and Christoph Studer$^1$} \\
\thanks{The work of OC was supported in part by ComSenTer, a Semiconductor Research Corporation (SRC) program, by SRC nCORE task 2758.004, and by a Qualcomm Innovation Fellowship. The work of SJ and GD was supported by the Swedish Foundation for Strategic Research under grant ID14-0022, and by the Swedish Governmental Agency for Innovation Systems (VINNOVA). The work of TG was supported in part by the US NSF under grant CCF-1535902 and by the US Office of Naval Research under grant N00014-17-1-2078. The work of CS was supported in part by Xilinx Inc.\ and by the US NSF under grants ECCS-1408006, CCF-1535897,  CCF-1652065, CNS-1717559, and ECCS-1824379.}\thanks{A MATLAB simulator for the FAWP approach proposed in this paper is available on GitHub: https://github.com/quantizedmassivemimo/fawp.}
\IEEEauthorblockA{\em $^1$Cornell Tech, New York, NY; e-mail: oc66@cornell.edu, studer@cornell.edu\\
$^2$Ericsson Research, Gothenburg, Sweden; e-mail: sven.jacobsson@ericsson.com\\
$^3$Chalmers University of Technology, Gothenburg, Sweden; e-mail: durisi@chalmers.se\\
$^4$University of Maryland, College Park, MD; e-mail: tomg@cs.umd.edu}
}

\maketitle

% ================================================================================
% ================================================================================
% ================================================================================
%
\begin{abstract}
Power consumption of multi-user (MU) precoding is a major concern in all-digital massive MU multiple-input multiple-output (MIMO) base-stations with hundreds of antenna elements operating at millimeter-wave (mmWave) frequencies. We propose to replace part of the linear Wiener filter (WF) precoding matrix by a finite-alphabet WF precoding (FAWP) matrix, which enables the use of low-precision hardware that consumes low power and area. To minimize the performance loss of our approach, we present methods that efficiently compute FAWP matrices that best mimic the WF precoder. Our results show that FAWP matrices approach infinite-precision error-rate and error-vector magnitude performance with only 3-bit precoding weights, even when operating in realistic mmWave channels. Hence, FAWP is a promising approach to substantially reduce power consumption and silicon area in all-digital mmWave massive MU-MIMO systems.
\end{abstract}
%
%% ================================================================================
%% ================================================================================
%% ================================================================================
%
\section{Introduction}
Next-generation wireless systems are expected to achieve unprecedentedly high data rates by combining the large bandwidths available at millimeter-wave (mmWave) frequencies~\cite{swindlehurst14a} with the high spectral efficiency provided by massive multi-user (MU) multiple-input multiple-output (MIMO)~\cite{larsson14a}.
Unfortunately, base-station (BS) architectures for MU-MIMO systems, with hundreds of antenna elements, operating at the extreme sampling rates needed for wideband mmWave communication require excessively high power consumption and complex digital circuitry.
To keep the power consumption within acceptable bounds, research has mostly focused on hybrid analog-digital architectures~\cite{roh2014millimeter,sadhu20177,alkhateeb14b}.
Such hybrid BS architectures are, however, limited in their beamforming capabilities~\cite{alkhateeb14b,bjornson2019massive,dutta2019case}, which leads to reduced spectral efficiency.
Per contra, all-digital BS architectures~\cite{mo15b,roth2017achievable,jacobsson17b} do not suffer from such limitations.
While it is natural to believe that all-digital solutions are more power-hungry than hybrid architectures, recent results show that---by reducing the data-converter resolution---the radio-frequency circuitry and data-converters in an all-digital BS (i) have similar power consumption as in a hybrid BS \cite{dutta2019case} and (ii) enable superior spectral efficiency~\cite{roth2017achievable}.
Besides these recent results, the power consumption and silicon area of baseband processing in all-digital BS architectures are largely~unexplored.

\vspace{-0.03cm}
\subsection{Finite-Alphabet Equalization}
In our recent paper~\cite{castaneda19fame}, we investigated the power consumption and silicon area required for spatial equalization in the mmWave MU-MIMO uplink, i.e., when the user equipments (UEs) transmit to the BS.
We considered a system with $16$ UEs and $256$ BS antenna elements operating at a sampling rate of $2$\,G\,vectors/s.
For such system, our implementation results in $28$\,nm CMOS technology showed that, even when considering data converters with only $7$ bits of resolution, 
a simple, single-tap linear equalizer already requires $28$\,W and $129\,\text{mm}^2$~\cite{castaneda19fame}.
For higher sampling rates or systems with more BS antenna elements, UEs, or taps,  power consumption and area will increase even further.
Hence, to reduce both power and silicon area, we proposed \emph{finite-alphabet equalization}~\cite{castaneda19fame}, which uses coarsely quantized numbers to represent the entries of the equalization matrices, while minimizing the post-equalization mean-square error (MSE).
In summary, we showed that finite-alphabet equalizers enable a reduction in power and area by a factor of $3.9\times$ and $5.8\times$, respectively, while offering competitive error-rate performance to conventional, high-resolution equalizers~\cite{castaneda19fame}.

\vspace{-0.03cm}
\subsection{Contributions}
Similar to the case of equalization in the uplink, the power consumption and silicon area of precoding in the all-digital mmWave MU-MIMO downlink (BS transmits to UEs) is expected to be a major bottleneck, as high-dimensional data has to be processed at extremely high rates.
In order to reduce power consumption and silicon area of the precoding operation, we apply the concept of finite-alphabet matrices used for linear spatial equalization in~\cite{castaneda19fame} to linear precoding.
We propose two finite-alphabet precoding schemes to compute Wiener filter (WF)-optimal matrices, which are matrices that best mimic
the linear WF precoder. 
To demonstrate the effectiveness of the framework that we call \emph{finite-alphabet WF precoding} (FAWP), we evaluate its performance in terms of uncoded bit error-rate~(BER) and error-vector magnitude (EVM) for i.i.d.\ Rayleigh fading, and for line-of-sight~(LoS) and non-LoS mmWave channels.

\subsection{Notation}
Uppercase and lowercase boldface letters denote matrices and column vectors, respectively. 
For a matrix $\bA$, its Hermitian transpose, Frobenius norm, real part and imaginary part are $\bA^H$, $\|\bA\|_{F}$, $\Re\{\bA\}$, and $\Im\{\bA\}$, respectively.
The $M\times M$ identity matrix is $\bI_M$.
For the vector $\bma$, its $k$th entry, $\ell_2$-norm, and entry-wise complex conjugate are $a_k$, $\vecnorm{\veca}_2$, and $\bma^*$, respectively. 
The $k$th standard basis vector is $\bme_k$.
The set $\reals_+$ contains the non-negative real numbers.
The signum function~$\text{sgn}(\cdot)$ is defined as~$\text{sgn}(a)=+1$ for~$a \in \reals_+$ and~$\text{sgn}(a)=-1$ for $a\not\in\reals_+$, and is applied entry-wise to vectors.
The expectation operator with respect to the random vector $\bmx$ is $\Ex{\bmx}{\cdot}$.
%

%%%
\section{System Model and WF Precoding}
\subsection{System Model}
We focus on the downlink of a mmWave massive MU-MIMO system in which a BS with~$B$ antennas serves~$U<B$ single-antenna UEs in the same time-frequency resource. 
We consider a narrowband scenario modeled by $\bmy = \bH \bmx + \bmn$,
where $\bmy\in\complexset^U$ is the received vector,
$\bH\in\complexset^{U\times B}$ is the channel matrix,
$\bmx\in\complexset^B$ is the precoded vector,
and $\bmn\in\complexset^U$ is i.i.d.\ circularly-symmetric complex Gaussian noise with variance~$\No$ per complex entry.
We assume that the channel matrix $\bH$ is perfectly known to the BS, and that the precoded vector $\bmx$ is subject to the following  average power constraint:
\begin{align}\label{eq:powerconstraint}
 \Ex{\bmx}{\|\bmx\|^2_2}\leq P.
\end{align}

\subsection{WF Precoding}
\label{sec:wfprecoding}

The goal of precoding is to simultaneously transmit constellation points $s_u\in\setO$ to the $u=1,\ldots,U$ UEs while reducing MU interference. 
Here, $s_u$ is assumed to have zero mean and variance $\Es$, and $\setO$ denotes the constellation set (e.g., $16$-QAM).
The BS maps the vector $\bms$ into the precoded vector $\bmx$ with the aid of channel state information. 
The precoded vector $\bmx$ is crafted such that the UEs can form an estimate $\hat s_u \in \opC$ of the transmitted symbol $s_u$ simply by scaling the received signal $y_u$.
Specifically, as in \cite{jacobsson17d,jacobsson16d}, we assume that each UE forms an estimate as $\hat s_u = \beta y_u$.
Here, $\beta\in\reals_+$ is a precoding factor that can be estimated at the UE using pilot-based transmission~\cite{jacobsson16d}.

In what follows, we focus on linear precoders for which it holds that $\bmx=\bP\bms$, where $\bP\in\complexset^{B\times U}$ is the precoding matrix.
While the literature covers a range of optimization criteria for linear precoding~\cite{fatema2017massive}, in this work we limit ourselves to the design of linear precoders that attempt to minimize the MSE between the estimated symbols $\hat\bms$ and the transmitted symbols~$\bms$:
\begin{align}
\textit{MSE} & = \Ex{\bms,\bmn}{\|\bms-\hat\bms\|^2_2} =  \Ex{\bms}{\|\bms-\beta\bH\bmx\|^2_2} + \beta^2 U \No. \label{eq:MSE}
\end{align} 
Minimizing~\fref{eq:MSE} subject to the power constraint in \fref{eq:powerconstraint} results in the so-called WF precoder~\cite{joham05a}, where the precoding matrix is given by $\bP^\text{WF}=\frac{1}{\beta^\text{WF}}\bQ^\text{WF}$ with 
\begin{align} \label{eq:originalQmatrix}
\bQ^\text{WF} = \left(\bH^H\bH + \kappa^\text{WF} \bI_B \right)^{-1}\bH^H,
\end{align}
\vspace{-0.6cm}
\begin{align} \label{eq:precodingfactor}
\kappa^\text{WF} =  \frac{U\No}{P}, \quad \text{and} \quad \beta^\text{WF}  = \sqrt{\frac{\tr\big((\bQ^\text{WF})^H\bQ^\text{WF}\big) \Es}{P}}.
\end{align}

It is important to realize that the matrix $\bQ^\text{WF} \in \opC^{B \times U}$ in \fref{eq:originalQmatrix} is the solution of the following optimization~problem:
\begin{align}\label{eq:preFAWPproblemMatrix} 
\bQ^\text{WF} = \argmin_{\tilde\bQ\in\complexset^{B\times U}} \|\bI_U-\bH\tilde\bQ\|_F^2 + \kappa^\text{WF} \|\tilde\bQ\|^2_F.
\end{align}
We can also obtain the columns~$\bmq^\text{WF}_u\in \opC^B$ , $u=1,\ldots,U$, of the matrix  $\bQ^\text{WF}$ by solving
\begin{align}\label{eq:preFAWPproblem} 
\bmq^\text{WF}_u  = \argmin_{\tilde\bmq\in\complexset^{B}} \|\bme_u-\bH\tilde\bmq\|_2^2 +  \kappa^\text{WF} \|\tilde\bmq\|^2_2.
\end{align}
By applying the Woodbury identity \cite{petersen12a} to \fref{eq:originalQmatrix}, we also have that
\begin{align} \label{eq:woodburyQmatrix}
\bQ^\text{WF} = \bH^H\left(\bH\bH^H + \kappa^\text{WF} \bI_U \right)^{-1},
\end{align}
which is the solution to the following optimization problem:
\begin{align}\label{eq:postFAWPproblemMatrix} 
\bQ^\text{WF} = \argmin_{\tilde\bQ\in\complexset^{B\times U}} \|\bI_B-\tilde\bQ\bH\|_F^2 + \kappa^\text{WF} \|\tilde\bQ\|^2_F.
\end{align}
Thus, the rows~$\bmq_b^\text{r,WF}$, $b=1,\ldots,B$, of $\bQ^\text{WF}$ (where the superscript $\text{r}$ denotes a row vector) can be computed as
\begin{align}\label{eq:postFAWPproblem} 
\bmq_b^\text{r,WF}=\argmin_{\tilde\bmq^\text{r}\in\complexset^{1\times U}} \|\bme_b^H-\tilde\bmq^\text{r}\bH\|_2^2 +  \kappa^\text{WF} \|\tilde\bmq^\text{r}\|^2_2.
\end{align}
The alternative optimization problems in \fref{eq:preFAWPproblem} and \fref{eq:postFAWPproblem} to compute the matrix $\bQ^\text{WF}$ will become useful in the next section.

\section{Finite-Alphabet WF Precoding~(FAWP)}

WF precoding computes $\bmx=\bP^\text{WF}\bms=\frac{1}{\beta^\text{WF}}\bQ^\text{WF}\bms$ for each transmitted vector $\bms$. 
Unfortunately, digital precoding circuitry will be power hungry and large as mmWave MU-MIMO systems operate with high-dimensional data and extremely high sampling rates.
As a remedy, FAWP proposes to represent the matrix $\bQ^\text{WF}$ using coarsely quantized numbers, with the objective of reducing the hardware complexity of the matrix-vector product $\bQ^\text{WF}\bms$.
Unfortunately, a direct quantization of the matrix $\bQ^\text{WF}$ typically leads to a significant error-rate degradation.

In order to design low-resolution matrices that are WF-optimal, i.e., that best mimic the infinite-precision WF-precoding matrix $\bQ^\text{WF}$, we propose to use the so-called \emph{finite-alphabet matrices}, initially proposed in~\cite{castaneda19fame} for spatial equalization in the mmWave MU-MIMO uplink.
Since we will apply finite-alphabet matrices to imitate the WF-precoding matrix $\bQ^\text{WF}$, we will refer to them as \emph{FAWP matrices}.
FAWP matrices introduce a \emph{few} high-resolution scaling factors that help to bring a low-resolution matrix to the right scale.
While the work in \cite{castaneda19fame} studied one form of finite-alphabet matrices, we will now consider two distinct FAWP matrix structures, namely pre-FAWP and post-FAWP matrices.

\subsection{Pre-FAWP Matrix}
\begin{defi}\label{def:preFAWPmatrix}
We define a \emph{pre-FAWP matrix} as a $B\times U$ matrix with the structure
\begin{align} \label{eq:preFAWPmatrix}
\bQ=\bA\,\mathrm{diag}(\boldsymbol\alpha^*),
\end{align}
where $\bA\in\setX^{B\times U}$ is a low-resolution matrix with entries taken from the finite alphabet $\setX$ and $\boldsymbol\alpha\in\complexset^U$ is a vector with per-UE scaling factors.
\end{defi}

By using a pre-FAWP matrix, the matrix-vector product $\bQ\bms$ becomes $\bA(\mathrm{diag}(\boldsymbol\alpha^*)\bms)$.
We call such matrix pre-FAWP as the~$U$ entries of the transmitted symbol vector $\bms$ are scaled by the entries of $\boldsymbol\alpha^*$ \emph{before} getting multiplied with the matrix $\bA$.
Pre-FAWP reduces hardware complexity of $\bQ\bms$ since the matrix~$\bA$ has low-resolution entries.
Consider, for example, the case in which the entries of~$\bA$ are chosen from the $1$-bit alphabet $\setX=\{\pm1\pm j\}$; multiplying this matrix $\bA$ with the vector $\mathrm{diag}(\boldsymbol\alpha^*)\bms$ does not require hardware multipliers, but only adders and subtractors.

To calculate pre-FAWP matrices that are WF-optimal, we solve the problem in~\fref{eq:preFAWPproblem} by assuming that $\bQ$ has the form given by~\fref{eq:preFAWPmatrix}.
By doing so, we arrive at the following procedure:
\begin{lem} \label{lem:preFAWPequivalent}
The problem in \fref{eq:preFAWPproblemMatrix} is equivalent to solving the following optimization problem for each UE $u=1,\ldots,U$:
\begin{align}\label{eq:preFAWPproblemCompact}
\bma_u = \argmin_{\tilde\bma\in\setX^B} \frac{\|\bH\tilde\bma\|_2^2+\kappa^\textnormal{WF}\|\tilde\bma\|_2^2}{|\bmh_u^\textnormal{r}\tilde\bma|^2 }.
\end{align}
Here, $\bma_u$ is the $u$th column of $\bA$, $\bmh_u^\textnormal{r}$ is the $u$th row of $\bH$, and the associated optimal scaling factor is given by
\begin{align}\label{eq:preFAWPscaling}
\alpha_u(\bma_u) = \frac{\bmh_u^\textnormal{r}\bma_u}{\|\bH\bma_u\|^2_2+\kappa^\textnormal{WF}\|\bma_u\|^2_2}.
\end{align}
\end{lem}
\fref{lem:preFAWPequivalent} can be established by first plugging \fref{eq:preFAWPmatrix} into \fref{eq:preFAWPproblem}.
Then, we obtain \fref{eq:preFAWPscaling} by taking the Wirtinger derivative with respect to $\alpha_u$.
Substituting \fref{eq:preFAWPscaling} in \fref{eq:preFAWPproblem} gives \fref{eq:preFAWPproblemCompact}; the proof is analogous to that in \cite{castaneda19fame} for finite-alphabet equalizers.

\subsection{Post-FAWP Matrix}
\begin{defi}\label{def:postFAWPmatrix}
We define a \emph{post-FAWP matrix} as a $B\times U$ matrix with the structure
\begin{align} \label{eq:postFAWPmatrix}
\bQ=\mathrm{diag}(\boldsymbol\zeta)\,\bZ^H,
\end{align}
where $\bZ\in\setX^{U\times B}$ is a low-resolution matrix with entries taken from the finite alphabet $\setX$ and $\boldsymbol\zeta\in\complexset^B$ is a vector with per-BS-antenna scaling factors.
\end{defi}

By using a post-FAWP matrix, the matrix-vector product $\bQ\bms$ becomes $\mathrm{diag}(\boldsymbol\zeta)(\bZ^H\bms)$.
We call such matrix post-FAWP as the $B$ scaling factors in $\boldsymbol\zeta$ are applied \emph{after} multiplying the matrix $\bZ^H$ with the vector $\bms$. 
Post-FAWP reduces the hardware complexity of $\bQ\bms$ since the $B\times U$ matrix-vector product $\bZ^H\bms$ can be implemented using exclusively low-resolution arithmetic units.
The results of $\bZ^H\bms$ are then entry-wise scaled by $\boldsymbol\zeta$, which requires only~$B$ high-resolution scalar multiplications.

Akin to the case of pre-FAWP matrices, we obtain post-FAWP matrices that are WF-optimal by solving the problem in \fref{eq:postFAWPproblem} with a matrix~$\bQ$ that has the form given in \fref{eq:postFAWPmatrix}.
By doing so, we arrive at the following procedure: 
\begin{lem} \label{lem:postFAWPequivalent}
The problem in \fref{eq:postFAWPproblemMatrix} is equivalent to solving the following optimization problem for each BS antenna $b=1,\ldots,B$:
\begin{align}\label{eq:postFAWPproblemCompact}
\bmz_b = \argmin_{\tilde\bmz\in\setX^U} \frac{\|\bH^H\tilde\bmz\|_2^2+\kappa^\textnormal{WF}\|\tilde\bmz\|_2^2}{|\bmh^H_b\tilde\bmz|^2 }.
\end{align}
Here, $\bmz_b$ is the $b$th column of $\bZ$, $\bmh_b$ is the $b$th column of $\bH$, and the associated optimal scaling factor is given by
\begin{align}\label{eq:postFAWPScaling}
\zeta_b(\bmz_b) = \frac{\bmh_b^H\bmz_b}{\|\bH^H\bmz_b\|^2_2+\kappa^\text{WF}\|\bmz_b\|^2_2}.
\end{align}
\end{lem}
\noindent The proof of \fref{lem:postFAWPequivalent} parallels  that of \fref{lem:preFAWPequivalent}.

In summary, both pre-FAWP and post-FAWP matrices are composed by a low-resolution matrix and a set of scaling factors.
The difference is that a pre-FAWP matrix applies its~$U$ scaling factors \emph{before} the multiplication with the low-resolution matrix, whereas a post-FAWP matrix applies its $B$ scaling factors \emph{after} matrix multiplication. 
As $B\gg U$ in typical massive MU-MIMO systems, a pre-FAWP matrix performs fewer high-resolution scaling operations than a post-FAWP matrix.
However, the matrix-vector product is simpler with a post-FAWP matrix than with a pre-FAWP matrix, since the vector has a lower resolution as the symbols in $\bms$ are not scaled yet.
Thus, neither pre-FAWP nor post-FAWP matrices have a clear advantage over the other in terms of hardware complexity.\footnote{In contrast, for the uplink considered in~\cite{castaneda19fame}, post-equalization scaling requires fewer scaling factors \emph{and} does not increase the resolution of the received~vector.}
Nonetheless, both FAWP matrix structures are expected to reduce hardware complexity when compared to traditional precoding, as the low-resolution matrices in both structures have \emph{coarsely quantized} entries.

\section{Computing FAWP Matrices}
We now propose different methods to compute pre-FAWP and post-FAWP matrices defined in \fref{eq:preFAWPmatrix} and \fref{eq:postFAWPmatrix}, respectively.
We also discuss means to estimate the precoding factor $\beta$.

\subsection{FAWP by Quantizing the WF-Precoding Matrix}
For pre-FAWP and post-FAWP matrices, the scaling factors are computed by means of \fref{eq:preFAWPscaling} and \fref{eq:postFAWPScaling}, respectively, regardless of how the low-resolution matrix ($\bA$ for pre-FAWP and $\bZ$ for post-FAWP) is computed.
Instead of solving the problems in \fref{eq:preFAWPproblemCompact} or \fref{eq:postFAWPproblemCompact}, a simple approach is to directly quantize the infinite-precision matrix $\bQ^\text{WF}$.
We call this approach \emph{FAWP-WF}; more specifically, pre-FAWP-WF and post-FAWP-WF when applied to pre-FAWP and post-FAWP matrices, respectively.

We quantize $\bQ^\text{WF}$ following the method put forward in \cite{castaneda19fame}. For pre-FAWP-WF,  we first find the maximum value $w_\text{max}$ of $[|\Re\{\bmq^\text{WF}_u\}|; |\Im\{\bmq^\text{WF}_u\}|]$ for each column $\bmq^\text{WF}_u$ of $\bQ^\text{WF}$.
We then divide the range $[-w_\text{max},w_\text{max}]$ into uniform-width bins, where each bin is represented by its centroid value.
The centroid values are scaled by the same factor so that they are integer numbers, which preserves the objective value in \fref{eq:preFAWPproblemCompact} and results in the low-resolution entries of the column~$\bma_u$.
For post-FAWP-WF, we apply the same procedure on a per-row basis: Each quantized row of $\bQ^\text{WF}$ corresponds to one row of~$\bZ^H$.

Since the problems in \fref{eq:preFAWPproblemCompact} and \fref{eq:postFAWPproblemCompact} are NP-hard, FAWP-WF significantly reduces complexity. 
Concretely, FAWP-WF requires the same complexity of $\setO(BU^2)$ as computing the infinite-precision $\bQ^\text{WF}$ in \fref{eq:woodburyQmatrix}.
As a result, we will use FAWP-WF as a baseline to evaluate the performance of the algorithm proposed next, which tackles the problems in \fref{eq:preFAWPproblemCompact} and \fref{eq:postFAWPproblemCompact}.

\subsection{FAWP via Forward-Backward Splitting (FBS)}
Similar to finite-alphabet equalization matrices in~\cite{castaneda19fame}, we can also approximately solve the FAWP problems in \fref{eq:preFAWPproblemCompact} and \fref{eq:postFAWPproblemCompact} using forward-backward splitting (FBS), an approach dubbed FAWP-FBS.
In what follows, we will present pre-FAWP-FBS, an algorithm for computing the low-resolution part of a pre-FAWP matrix starting from the problem in \fref{eq:preFAWPproblemCompact}. 
The algorithm for post-FAWP matrices, dubbed post-FAWP-FBS, can be derived in a similar way starting from \fref{eq:postFAWPproblemCompact}.

As in \cite{castaneda19fame}, we assume that the optimal objective value $\gamma_u$ of~\fref{eq:preFAWPproblemCompact}, $u=1,\ldots,U$, is known.
Then, solving the problem in~\fref{eq:preFAWPproblemCompact} is equivalent to solving the following problem:
\begin{align}\label{eq:FAWP2solve}
\bma_u = \argmin_{\tilde\bma\in\setX^B} \frac{1}{2}\|\bH\tilde\bma\|_2^2+\frac{\kappa^\text{WF}}{2}\|\tilde\bma\|_2^2 - \frac{\gamma_u}{2}|\bmh_u^\text{r}\tilde\bma|^2.
\end{align}
As $\gamma_u$ is unknown, we will use it as a parameter that can be tuned to empirically improve the performance of our algorithm.

We next relax the finite-alphabet constraint $\tilde\bma\in\setX^B$ in \fref{eq:FAWP2solve} to $\tilde\bma\in\setB^B$, where $\setB$ represents the convex hull of $\setX$.
By doing so, the all-zeros vector $\bm0_{B\times1}$ becomes a trivial solution.
To avoid this solution, we follow the approach in \cite{shah2016biconvex} and include in \fref{eq:FAWP2solve} the term $-\frac{\delta}{2}\|\tilde\bma\|_2^2$, with $\delta>0$, to encourage large entries in the vector~$\tilde\bma$.
The resulting optimization problem is 
\begin{align}
\label{eq:FAWP_BCR}
\bma_u=\argmin_{\tilde{\bma}\in\setB^B}\frac{1}{2}\|\bH\tilde\bma\|_2^2-\frac{\gamma_u}{2}|\bmh^\text{r}_u\tilde\bma|^2+\frac{\kappa^\text{WF}-\delta}{2}\|\tilde\bma\|_2^2.
\end{align}

We are now ready to apply FBS~\cite{goldstein2010high,goldstein14a}.
FBS is an efficient procedure for solving convex problems of the form $\hat{\bma} = \argmin_{\tilde\bma} f(\tilde\bma) + g(\tilde\bma)$, where both functions $f$ and $g$ are convex, but $f$ is smooth and $g$ is not necessarily smooth or bounded.  
FBS is an iterative method that runs for $t_\text{max}$ iterations or until convergence~\cite{goldstein14a}. In each iteration $t$, FBS computes
\begin{align}
\tilde\bmv^{(t+1)} & =\tilde\bma^{(t)}-\tau^{(t)}\nabla f(\tilde\bma^{(t)}) \label{eq:fbs_z},\\
\tilde\bma^{(t+1)} & =\text{prox}_g\left(\tilde\bmv^{(t+1)};\tau^{(t)}\right)\!,
\end{align}
where $\nabla f(\tilde\bma^{(t)})$ is the gradient of the function $f$ and $\lbrace\tau^{(t)}>0\rbrace$ is a sequence of step sizes.
The proximal operator of the function $g$ is defined as $\text{prox}_g\left(\tilde\bmv;\tau\right) = \argmin_{\tilde\bma} \left\lbrace \tau g(\tilde\bma) + \frac{1}{2}\|\tilde\bma-\tilde\bmv\|^2_2\right\rbrace$~\cite{parikh14a}.

Since the problem in \fref{eq:FAWP_BCR} is non-convex, FBS is not guaranteed to converge to an optimal solution.
Nevertheless, we use FBS to approximately solve \fref{eq:FAWP_BCR} by setting 
\begin{align}
f(\tilde\bma) & = \frac{1}{2}\|\bH\tilde\bma\|_2^2-\frac{\gamma_u}{2}|\bmh^\text{r}_u\tilde\bma|^2,\\
g(\tilde\bma) & =  \mathbb{I}_{\setB^B}(\tilde\bma)+\frac{\kappa^\text{WF}-\delta}{2}\|\tilde\bma\|^2_2,
\end{align}
where $\mathbb{I}_{\setB^B}(\tilde\bma)$ is the indicator function, which is zero if $\tilde\bma\in\setB^B$ and infinity otherwise.
We use the indicator function to incorporate the convex constraint $\tilde{\bma}\in\setB^B$ in \fref{eq:FAWP_BCR} into the function $g(\tilde\bma)$.
These choices for $f(\tilde\bma)$ and $g(\tilde\bma)$ result in:
\begin{align}
\nabla f(\tilde\bma) = &~\bH^H\bH\tilde\bma-\gamma_u(\bmh^\text{r}_u)^H\bmh^\text{r}_u\tilde\bma \label{eq:fbs_grad}\\
\text{prox}_g\left(\tilde{v}\right) = &~\text{sgn}\lefto(\Re\lbrace\tilde{v}\rbrace\right)\min\left\{\nu^{(t)}|\Re\lbrace\tilde{v}\rbrace|,1\right\} \nonumber\\
& + j~\text{sgn}\lefto(\Im\lbrace\tilde{v}\rbrace\right)\min\left\{\nu^{(t)}|\Im\lbrace\tilde{v}\rbrace|,1\right\} \label{eq:proxgFAME},
\end{align}
where $\nu^{(t)}=(1+\tau^{(t)}(\kappa^\text{WF}-\delta))^{-1}$ and \fref{eq:proxgFAME} is applied element-wise to $\tilde\bmv$.
Pre-FAWP-FBS can be summarized as follows:

\begin{oframed}
\vspace{-0.25cm}
\begin{alg}[Pre-FAWP-FBS]\label{alg:pre-FAWP-FBS} 
Initialize $\tilde\bma^{(1)}$ with either the maximum-ratio transmission (MRT) solution $(\bmh^\textnormal{r}_u)^H$ or the pre-FAWP-WF solution $\bma^\textnormal{WF}_u$, and fix the sets of parameters {$\{\tau^{(t)}\}$}, {$\{\nu^{(t)}\}$}, and {$\{\gamma^{(t)}\}$}.
Then, for each iteration $t=1,2,\ldots,t_\textnormal{max}$, compute
\begin{align}
\tilde\bmv^{(t+1)} &= \left(\bI_B - \tau^{(t)}\bH^H(\bI_U-\gamma^{(t)}\bme_u\bme_u^H)\bH\right) \tilde\bma^{(t)} \label{eq:step1}\\
\tilde\bma^{(t+1)} & = \mathrm{prox}_g (\tilde\bmv^{(t+1)}). \label{eq:step2}
\end{align}
The result $\tilde\bma^{(t_\text{max}+1)}$ is projected onto the finite alphabet $\setX$ to obtain $\bma_u$.
The optimal scalar $\alpha_u$ is computed using \fref{eq:preFAWPscaling}.
This procedure is repeated for each UE $u=1,\ldots,U$.
\end{alg}
\vspace{-0.25cm}
\end{oframed}
\vspace{-0.1cm}

To tune the algorithm parameters $\{\tau^{(t)}\}$, $\{\nu^{(t)}\}$, and $\{\gamma^{(t)}\}$, we use a neural-network-based approach as put forward in~\cite{balatsoukas-stimming19a}.
Note that we have replaced $\gamma_u$ with $\gamma^{(t)}$ in \fref{alg:pre-FAWP-FBS} in order to (i) keep the algorithm general for different user locations and (ii) to increase flexibility during optimization.

We now summarize post-FAWP-FBS, which can be derived following similar steps as for the derivation of pre-FAWP-FBS.
\begin{oframed}
\vspace{-0.25cm}
\begin{alg}[Post-FAWP-FBS]\label{alg:post-FAWP-FBS} 
Initialize $\tilde\bmz^{(1)}$ with either the MRT solution $\bmh_b$ or the post-FAWP-WF solution $\bmz^\text{WF}_b$, and fix the sets of parameters {$\{\tau^{(t)}\}$}, {$\{\nu^{(t)}\}$}, and {$\{\gamma^{(t)}\}$}.
Then, for each iteration $t=1,2,\ldots,t_\text{max}$, compute 
\begin{align}
\tilde\bmv^{(t+1)} &= \left(\bI_U - \tau^{(t)}\bH(\bI_B-\gamma^{(t)}\bme_b\bme_b^H)\bH^H\right) \tilde\bmz^{(t)} \label{eq:step1}\\
\tilde\bmz^{(t+1)} & = \mathrm{prox}_g (\tilde\bmv^{(t+1)}). \label{eq:step2}
\end{align}
The result $\tilde\bmz^{(t_\text{max}+1)}$ is projected onto the finite alphabet $\setX$ to obtain $\bmz_b$.
The optimal scale $\zeta_u$ is computed with \fref{eq:postFAWPScaling}.
This procedure is done for each BS antenna $b=1,\ldots,B$.
\end{alg}
\vspace{-0.25cm}
\end{oframed}
\vspace{-0.1cm}

We note that both FAWP-FBS algorithms have the same complexity order of $\setO(BU^2)$ as WF and FAWP-WF.

\subsection{Estimating the Precoding Factor $\beta$}
\label{sec:est_beta}
While the BS is  able to compute the precoding factor $\beta$ via~\fref{eq:precodingfactor} with a FAWP matrix $\bQ$ instead of $\bQ^\text{WF}$, the UEs need to estimate such precoding factor in order to correctly estimate the transmitted symbols in $\bms$.
As shown in \cite{jacobsson16d}, estimation can be achieved in a block-fading scenario by transmitting a pilot symbol that is known at the UE side.
Specifically, the BS will transmit the pilot $s_u=\sqrt{\Es},~u=1,\ldots,U$.
Then, the $u$th UE will receive ${y}_u=\beta^{-1}\bmh^\text{r}_u\bmq_u {s}_u+\check{e}_u+n_u$, where $\check{e}_u$ represents residual interference from the other UEs.
The objective now is for the UE to find a $\hat\beta_u\in\reals_+$ such that it generates an unbiased estimate $\hat{s}_u$ of $s_u$, i.e., $\hat{s}_u=\hat\beta_u y_u \approx s_u$.
By taking into account that the transmitted pilot symbol $s_u$ is known to be $\sqrt{\Es}$ and by assuming that $\check{e}_u+n_u$ is zero-mean Gaussian distributed and independent of $s_u$, the UE can compute a maximum likelihood estimate (MLE) of $\hat\beta_u$ as \cite{jacobsson16d}:
\begin{align}\label{eq:betamle}
{\hat\beta}^\text{MLE}_u=\reals\{\sqrt{\Es}/y_u\}.
\end{align}
While more pilots could be transmitted to form a better estimate ${\hat\beta}^\text{MLE}_u$, our results in \fref{sec:results} show that one pilot is sufficient to achieve reliable downlink communication.

\section{Numerical Results}\label{sec:results}

We now present simulation results for both pre-FAWP and post-FAWP matrices generated by either FAWP-WF or FAWP-FBS.
We perform a comparison in terms of BER and EVM versus normalized transmit power, which we define as $P/\No$.
For simplicity, we restrict our evaluation on a mmWave system with $B=256$ BS antennas serving $U=16$ UEs.

\subsection{1-bit FAWP BER Performance and $\beta$-Estimation}
\begin{figure}[tp]
\centering
\subfigure[Perfect  knowledge of $\beta$.]{\includegraphics[width=.48\columnwidth]{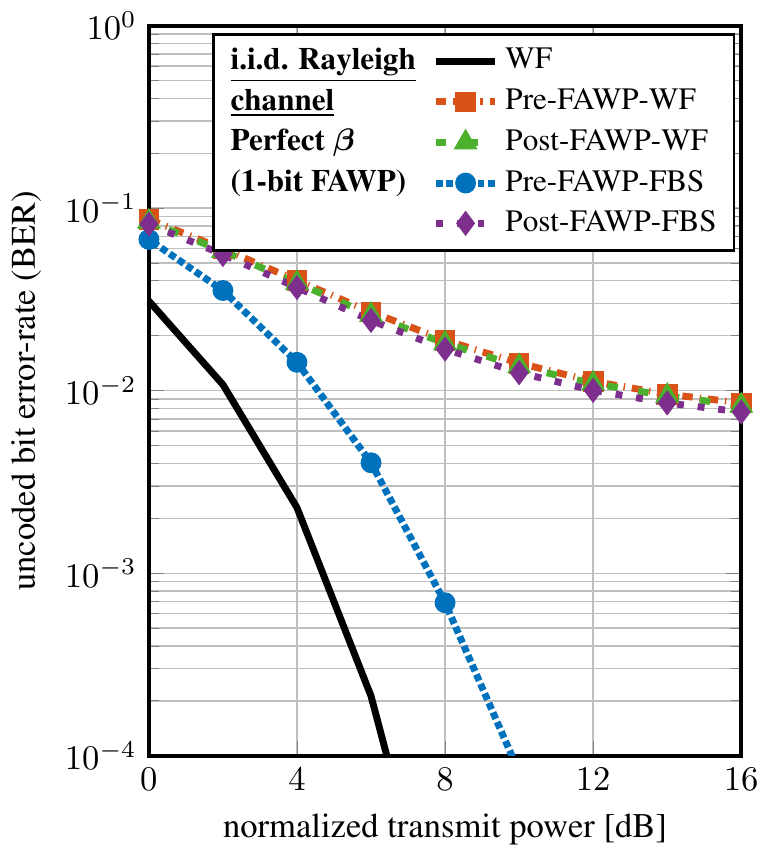}\label{fig:beta_perfect}}
\hfill
\subfigure[Estimated $\beta$ using one pilot.]{\includegraphics[width=.48\columnwidth]{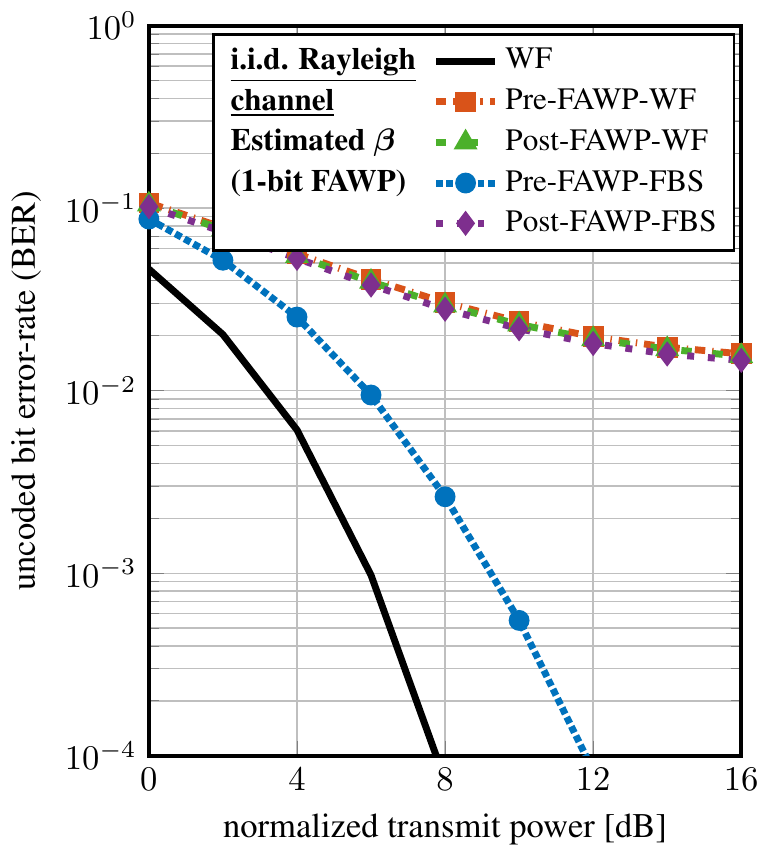}\label{fig:beta_estimated}}
\caption{Uncoded bit-error rate (BER) for a $B=256$ BS-antenna, $U=16$ UE, $16$-QAM system operating over an i.i.d. Rayleigh fading channel. All the FAWP-based approaches use $1$-bit FAWP matrices. Pre- and post-FAWP-FBS run for $t_\text{max}=10$ iterations starting from the MRT solution $\bH^H$.}\label{fig:beta_all}
\vspace{-0.3cm}
\end{figure}
\begin{figure*}[tp]
\centering
\subfigure[$1$-bit FAWP.]{\includegraphics[width=.64\columnwidth]{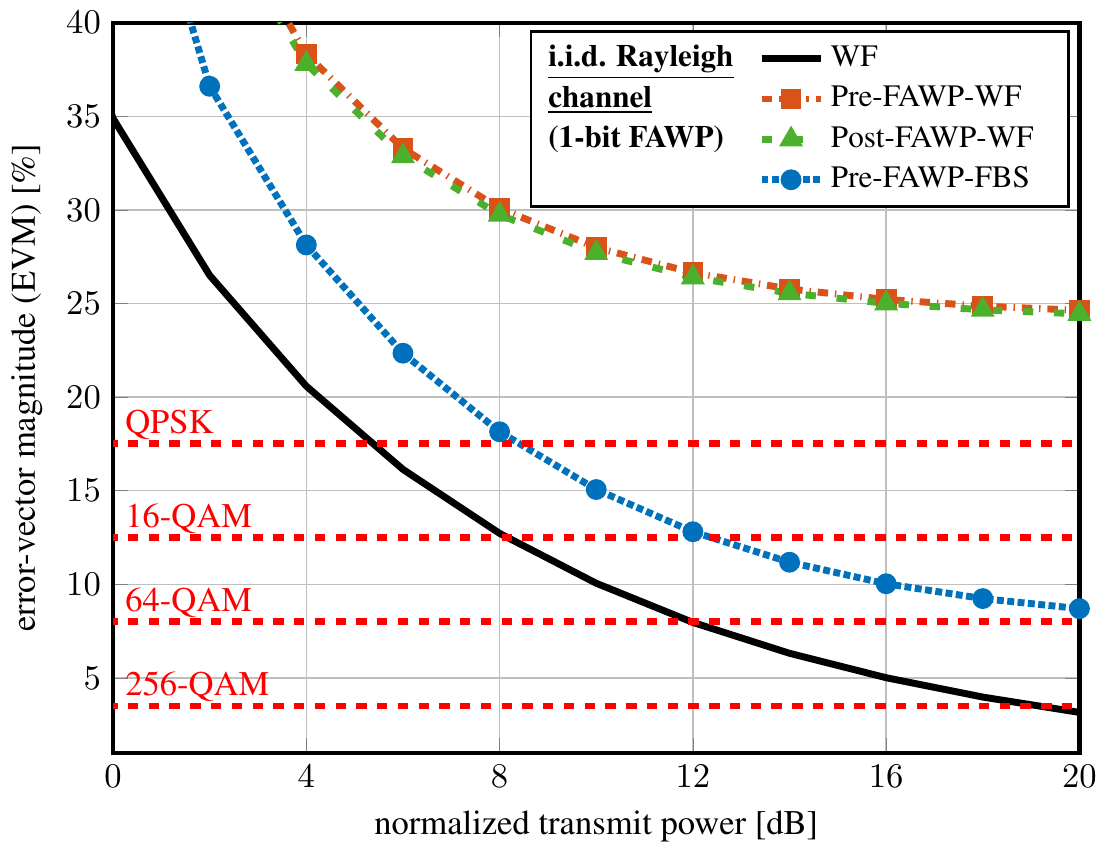}\label{fig:evm_1b}}
\hfill
\subfigure[$2$-bit FAWP.]{\includegraphics[width=.64\columnwidth]{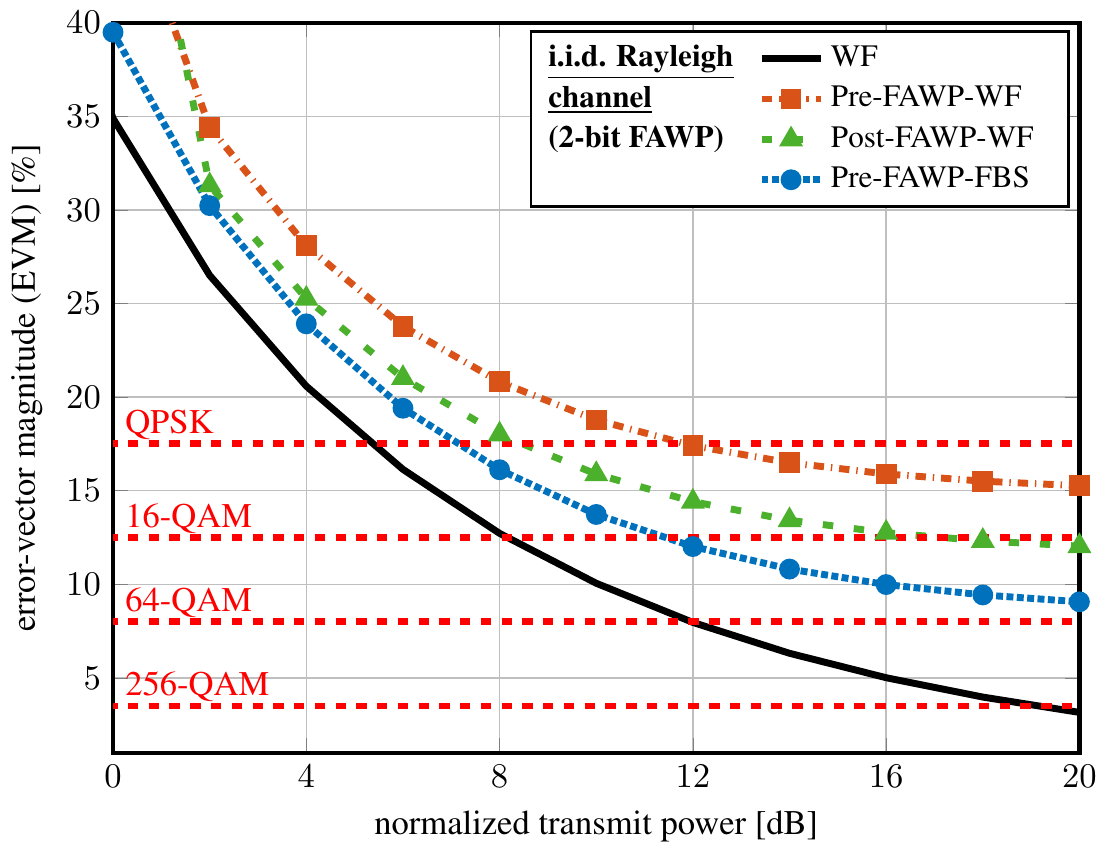}\label{fig:evm_2b}}
\hfill
\subfigure[$3$-bit FAWP.]{\includegraphics[width=.64\columnwidth]{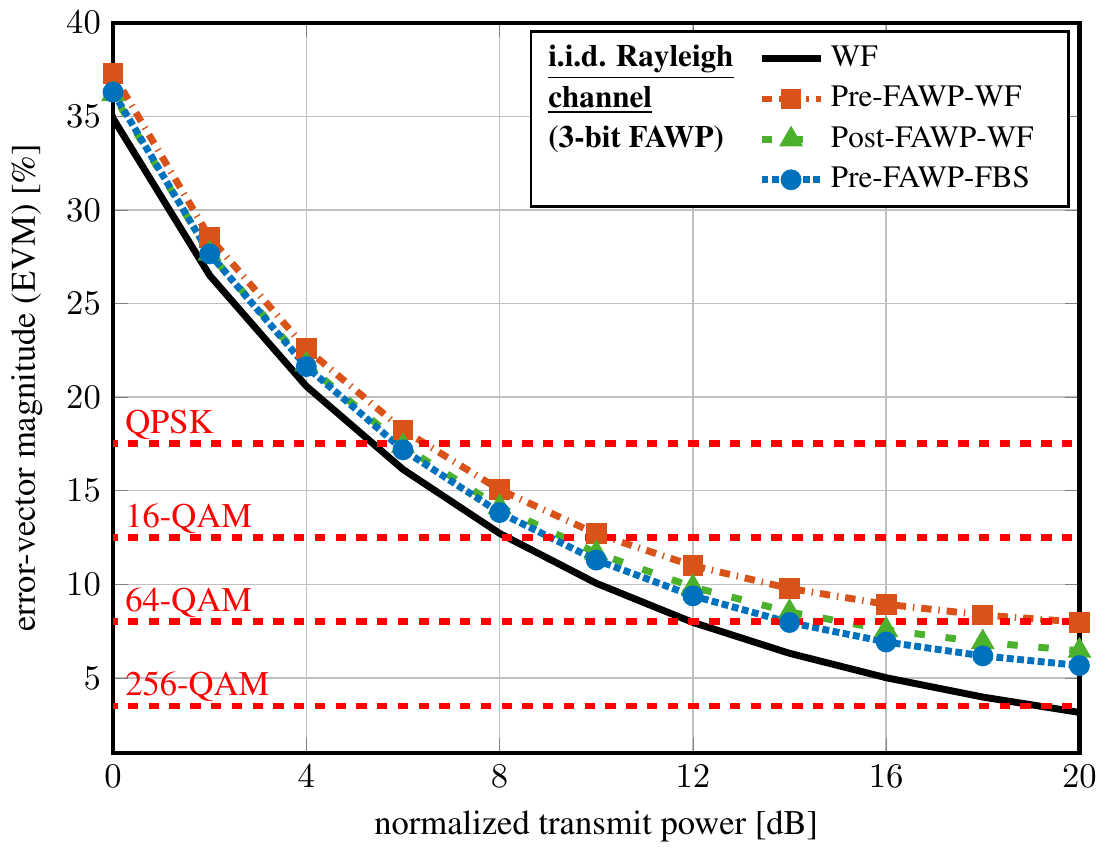}\label{fig:evm_3b}}
\caption{Error-vector magnitude (EVM) for a $B=256$ BS-antenna, $U=16$ UE system operating in an i.i.d. Rayleigh fading channel. The red dashed lines represent the EVM requirements established by the 3GPP 5G NR technical specification \cite{3gpp19a}. We consider FAWP using $\{1,2,3\}$-bit alphabets. For $\{2,3\}$-bit FAWP, pre-FAWP-FBS is initialized with the MRT solution $\bH^H$ and runs for $t_\text{max}=5$ iterations. The details for $1$-bit pre-FAWP-FBS are given in \fref{fig:beta_all}.}\label{fig:evm_all}
\vspace{-0.2cm}
\end{figure*}
\fref{fig:beta_all} shows the uncoded BER for the considered system when using $16$-QAM in an i.i.d. Rayleigh fading channel.
For the FAWP-based approaches, we use a $1$-bit alphabet.
In \fref{fig:beta_perfect}, we consider the case where the UEs have genie-aided access to the exact $\hat\beta_u$ precoding scaling factor.
Here, we can see that both FAWP-WF approaches result in an error floor.
In fact, the FAWP-WF precoders are significantly outperformed by pre-FAWP-FBS, which computes WF-optimal pre-FAWP matrices.
However, post-FAWP-FBS is unable to outperform post-FAWP-WF, a surprising behavior that we observe consistently across all our experiments---a detailed investigation of this behavior is left for future work. 
Hence, we exclude post-FAWP-FBS in the ensuing discussion.

In \fref{fig:beta_estimated}, we consider the same scenario as before, but this time $\hat\beta_u$ is estimated from a single pilot transmission as described in \fref{sec:est_beta}.
We can see that all precoders (including the infinite-precision WF) suffer from roughly a $2$\,dB loss.
In what follows, we assume that $\hat\beta_u$ is estimated using a single pilot. 

\subsection{Multi-Bit FAWP EVM Performance}
\fref{fig:evm_all} shows the EVM performance for the different FAWP precoders and $\{1,2,3\}$-bit alphabets.
The red dashed lines represent the per-modulation EVM requirements as specified by the 3GPP 5G NR standard~\cite{3gpp19a}.
\fref{fig:evm_1b} confirms what we previously observed in \fref{fig:beta_all} for the $1$-bit alphabet:
While FAWP-WF suffers a high error-floor that prevents such approach from reaching the EVM requirement even for QPSK, pre-FAWP-FBS almost meets the EVM requirement for $64$-QAM.
By increasing the number of bits used for the finite alphabet, the gap between the FAWP approaches and the infinite-precision WF decreases---to the point shown in \fref{fig:evm_3b} where all FAWP approaches meet the $64$-QAM EVM requirement when using a $3$-bit alphabet.
It is interesting to observe that post-FAWP-WF outperforms pre-FAWP-WF when using finite alphabets with more than $1$ bit.
Nonetheless, post-FAWP-WF is unable to outperform \mbox{pre-FAWP-FBS}.

\begin{figure}[tp]
\centering
\subfigure[QuaDRiGa non-LoS.]{\includegraphics[width=.48\columnwidth]{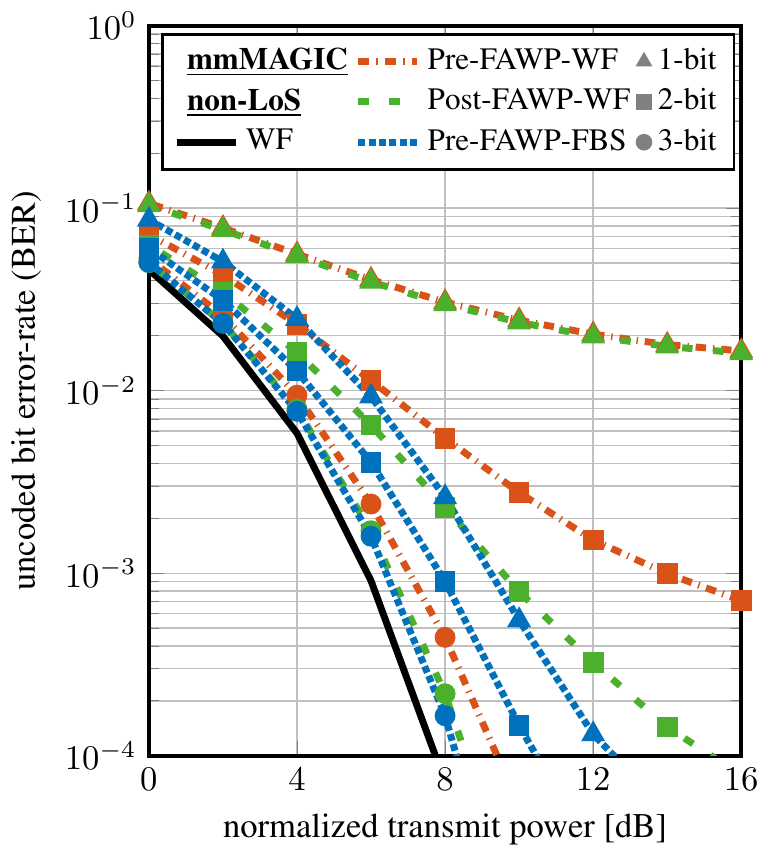}\label{fig:quadriga_nlos}}
\hfill
\subfigure[QuaDRiGa LoS.]{\includegraphics[width=.48\columnwidth]{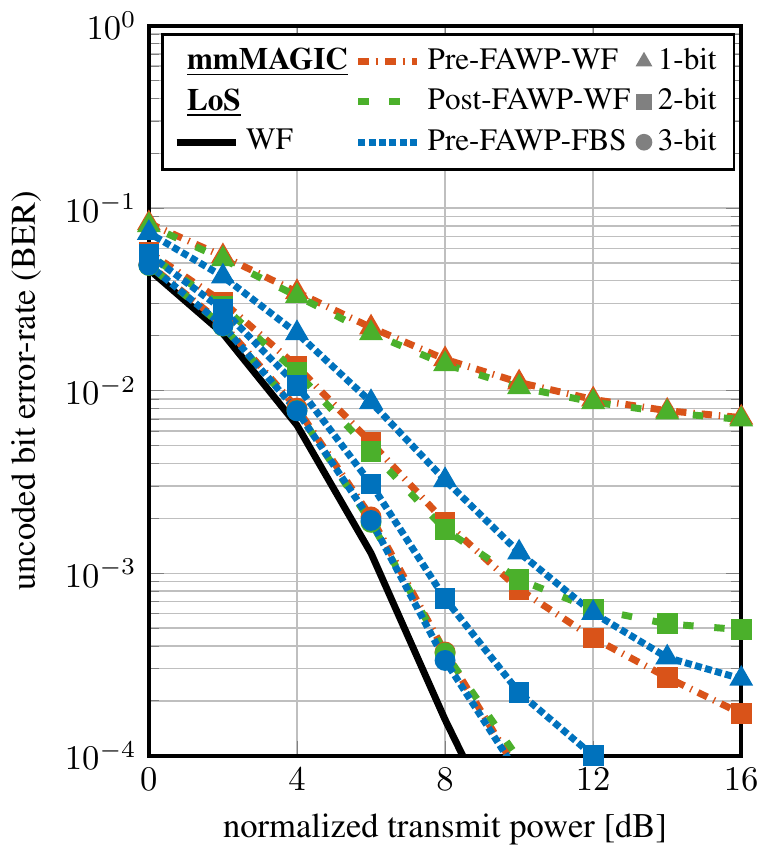}\label{fig:quadriga_los}}
\caption{Uncoded bit-error rate (BER) for a $B=256$ BS-antenna, $U=16$ UE, $16$-QAM system operating in realistic mmWave channel models. $1$-bit pre-FAWP-FBS runs $t_\text{max}=10$ iterations; $\{2,3\}$-bit pre-FAWP-FBS run no more than $t_\text{max}=5$ iterations. Pre-FAWP-FBS is initialized with $\bH^H$ for all cases but the $\{2,3\}$-bit LoS ones, which use $\bA^\text{WF}$ from pre-FAWP-WF.}\label{fig:quadriga_all}
\vspace{-0.2cm}
\end{figure}

\subsection{Performance Under Realistic Propagation Conditions}
We now evaluate FAWP under more realistic mmWave propagation conditions.
We use the QuaDRiGa channel model \cite{jaeckel2014quadriga} to simulate communication in the ``mmMAGIC\_UMi'' scenario when using a $60$\,GHz carrier frequency for both non-LoS and LoS propagation conditions.
We randomly place the UEs $10$\,m to $110$\,m away from the BS in a $120^\circ$ circular sector, with a minimum angular separation of $4^\circ$.
Furthermore, we assume perfect power control, i.e., all the users receive the same signal power.
Our simulation results are shown in \fref{fig:quadriga_all}.
The trends  we observed in the i.i.d.\ Rayleigh fading scenario are confirmed:
Pre-FAWP-FBS is able to outperform both FAWP-WF precoders, although the gains of the former (as well as the gap to the WF precoder) reduce when using more bits for the finite alphabet.
An interesting observation is that, for the LoS scenario illustrated in \fref{fig:quadriga_los}, the performance of pre-FAWP-WF is on par with that of post-FAWP-WF, which was not the case for the non-LoS and i.i.d.\ Rayleigh fading scenarios.
Regardless, the results in \fref{fig:quadriga_all} demonstrate that FAWP remains to perform well with realistic mmWave channels, which holds the promise of FAWP enabling low-power and area-efficient precoding~circuitry.

\section{Conclusions}
To enable energy- and area-efficient circuitry, we have proposed FAWP, an approach that replaces part of the linear WF precoder with a low-resolution matrix.
We have developed two structures for FAWP matrices, pre-FAWP and post-FAWP, as well as two methods to craft such matrices.
Our simulation results have shown that the sophisticated pre-FAWP-FBS algorithm is able to significantly outperform a simple quantization of the WF-precoding matrix, especially when using extremely low-resolution alphabets, and that it approaches the performance of the infinite-precision WF precoder with as few as $3$ bits of resolution.
Pre-FAWP-FBS accomplishes such feats while exhibiting the same asymptotical complexity as the WF precoder.
As for post-FAWP matrices, our simulation results have shown that post-FAWP-FBS does not outperform the simple quantization of the WF precoder.
We have verified these results under realistic conditions, such as LoS and non-LoS mmWave channels, as well as with estimation of the precoding factor $\beta$.
Thus, FAWP matrices are a promising approach to reduce hardware complexity and power consumption of precoding in mmWave MU-MIMO systems.
However,  in order to quantify the real-world benefits of FAWP, a hardware-level evaluation is necessary---such an evaluation is part of ongoing work. 
Since our FAWP approach performs matrix-vector products with coarsely quantized numbers, corresponding hardware implementations could benefit from emerging processing-in-memory architectures, such as the one proposed in~\cite{castaneda2019ppac}.

%
%% References should be produced using the bibtex program from suitable
%% BiBTeX files (here: strings, refs, manuals). The IEEEbib.bst bibliography
%% style file from IEEE produces unsorted bibliography list.
%% -------------------------------------------------------------------------
\balance

\bibliographystyle{IEEEtran}
\bibliography{IEEEabrv,confs-jrnls,publishers,vipbib}

\balance

\end{document}